# Electronic and Structural Characteristics of Zinc-Blende Wurtzite Biphasic Homostructure GaN Nanowires

Benjamin W. Jacobs,[1] Virginia M. Ayres,[1] Mihail P. Petkov,[2] Joshua B. Halpern,[3] MaoQe He,[4]
Andrew D. Baczewski,[1] Kaylee McElroy,[1] Martin A. Crimp,[5] Jiaming Zhang,[5] Harry C. Shaw[6]

[1]Department of Electrical and Computer Engineering, Michigan State University, East Lansing, Michigan 48824
[2]Jet Propulsion Laboratory, Pasadena, California 91109
[3]Department of Chemistry, Howard University, Washington, D.C. 20059
[4]Materials Science Research Center of Excellence, Howard University, Washington, D.C. 20059
[5]Department of Chemical Engineering and Materials Science, Michigan State University, East Lansing, Michigan 48824
[6]NASA Goddard Space Flight Center, Greenbelt, Maryland 20771

We report a new biphasic crystalline wurtzite/zinc-blende homostructure in gallium nitride nanowires. Cathodoluminescence was used to quantitatively measure the wurtzite and zinc-blende band gaps. High resolution transmission electron microscopy was used to identify distinct wurtzite and zinc-blende crystalline phases within single nanowires through the use of selected area electron diffraction, electron dispersive spectroscopy, electron energy loss spectroscopy, and fast Fourier transform techniques. A mechanism for growth is identified.

PACS numbers:

Over the past decade nanowires made from a wide variety of materials have demonstrated excellent electronic, mechanical, and chemical characteristics. Gallium nitride (GaN) nanowires in particular have shown potential for a wide range of optical and electronic applications. Room temperature UV lasing has been reported for GaN nanowire systems[1]. GaN nanowire field effect transistors[2] and logic devices[3] have shown desirable characteristics such as high transconductance and good switching. Various methods have been employed for growing GaN nanowires including chemical vapor deposition (CVD)[4], metal-organic CVD (MOCVD)[5], growth within a carbon nanotube[6] and laser ablation[7]. A metal catalyst is commonly used to initiate growth, and different growth directions have been achieved with different catalyst particles[8]. Catalyst free GaN nanowire growth via molecular beam expitaxy has also been recently reported, and employs a similar growth mechanism described here[9].

In these experiments, GaN nanowires are grown by a catalyst free, vapor-liquid process from the direct reaction of Ga metal vapor with $NH_3$ in a tube furnace[10,11]. At first, an amorphous, GaN matrix forms, followed by the appearance of small hexagonal platelets. By varying the ammonia flow rate (20-150sccm) and temperature (800-1100°C) the onset of nanowire growth and the diameters of the wires, which range between 10nm and 10μm, can be controlled. The length of the nanowires is determined by the duration of the growth cycle and can be millimeters in length. The nanowires form a single seamless structure along well defined crystal axes. A study of the temperature-composition parameter space provides a predictive model for nanowire growth[12].

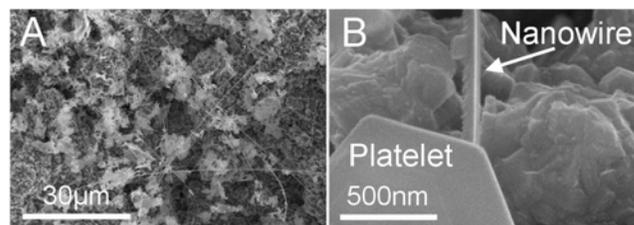

FIG 1. A. SEM image of the distribution of nanowires on the growth matrix. B. SEM image of the nanowire growing from a hexagonal platelet.

Scanning electron microscopy (Hitachi S-4700-II FESEM) images of the as grown GaN nanowire/amorphous/platelet matrix and the nanowire from platelet growth are shown in Figure 1(A,B). A relatively high density of nanowires is typically observed on the matrix surface.

In preparing the samples for cathodoluminescence and high resolution transmission electron microscopy, the GaN nanowire/amorphous matrix was ultrasonicated in ethanol for 5 seconds to separate the nanowires from the amorphous matrix. The mixture was then separated by centrifuging the solution at ~1000rpm for 30 seconds and decanting, thereby eliminating most of the amorphous matrix. For the cathodoluminescence analysis, nanowires were dispersed on a silicon wafer with a layer of native oxide. For the HRTEM analysis, a small droplet of the ethanol/nanowire solution was placed on a copper TEM grid with either a carbon holey film or carbon lacy film spanning the holes of the grid and allowed to dry.

Cathodoluminescence (CL) experiments were carried out on a LEO Supra 50 VP scanning electron microscope (SEM) in conjunction with a Gatan MonoCL system with a low-noise Peltier-cooled Hamamatsu R5509 photomultiplier detector. The GaN signature in the 290-360nm wavelength range was used to optimize the CL signal. The best peak-to-background ratio was achieved at approximately 10keV incident electron energy. The slit widths were set at 1mm, which was the optimum for achieving high count rates while maintaining sufficient resolution to differentiate between the wurtzite and zinc-blende band gaps; narrower slits did not change the peak widths significantly. In these initial investigations, detailed analysis of the near-band-gap exciton electronic structure was not considered and the experiments were carried out at room temperature.

The CL spectra of various GaN nanowires exhibited either a single peak, ascribed to the wurtzite structure, or two distinct peaks, implying the simultaneous presence of both wurtzite and zinc-blende structures. It should be noted that the spectra had virtually no notable contribution from luminescence that could be attributed to defects (stacking faults or impurity-mediated), which indicated high quality samples. The CL spectrum shown in Figure 2 shows evidence of the coexistence of zinc-blende and wurtzite structures in a single isolated GaN nanowire. The peak maxima, observed at ~3.64eV and ~3.88eV, are identified as the zinc-blende and wurtzite peaks based on their similarity to those observed for bulk GaN. The energy band gap values for bulk GaN are 3.2eV in bulk zinc-blende[13] and 3.39eV in bulk wurtzite[14]. The increase in energy of 0.45–0.5eV, measured by CL, may be attributed to both electron confinement and strain effects as discussed below.

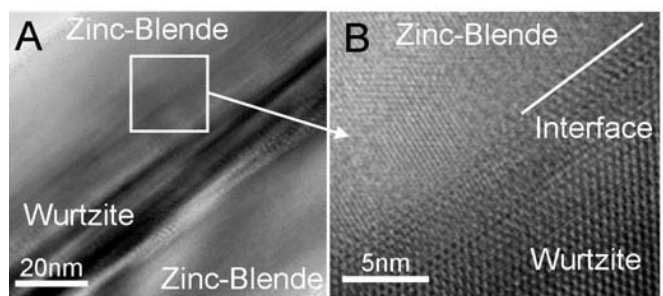

FIG 3. A. TEM image of GaN nanowire showing the inner wurtzite phase, dark contrast, and the outer zinc-blende phase, lighter contrast areas surrounding. The white box indicates where the close up in image B was taken. B. HRTEM image of the wurtzite-zinc-blende interface. The white line highlights where the phase transition occurs. This image also shows the highly crystalline nature of both phases.

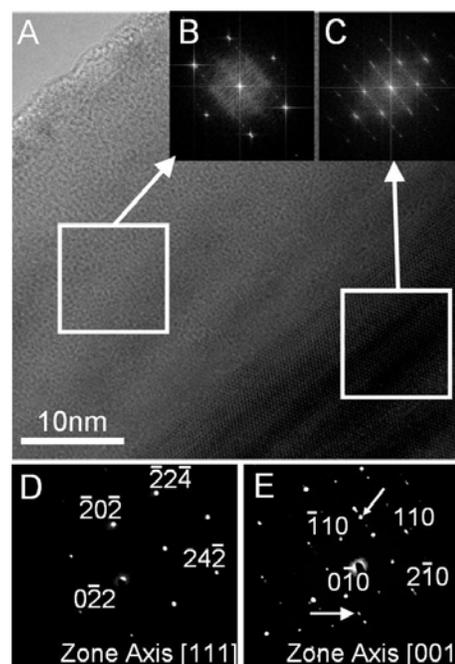

FIG 4. A. HRTEM image indicating where the FFTs and diffraction patterns were taken, shown by the white boxes. B. The FFT of the indicated area from A. C. FFT of the indicated area from A. D. Indexed diffraction pattern for the zinc-blende phase. The growth direction is [011]. E. Indexed diffraction pattern for the wurtzite phase. The growth direction is [110]. Arrows indicate diffraction contributing from the zinc-blende phase.

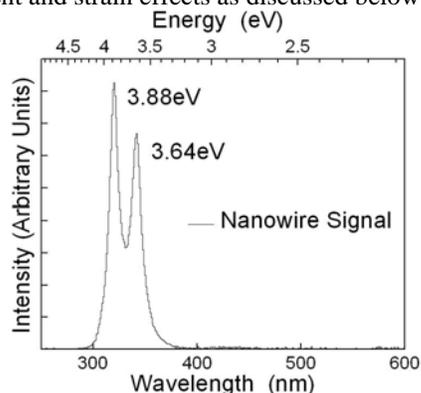

FIG 2. Cathodoluminescence spectrum showing the double peak indicating the wurtzite and zinc-blende phases present in the nanowire. The band gap energies for each phase is blue shifted on the order of 0.45-0.5eV.

The nanowire structure was investigated with high resolution transmission electron microscopy (HRTEM) using a JEOL 2200FS at the Center for Advanced Microscopy at Michigan State University. Selected area electron diffraction (SAED), and fast Fourier transforms (FFTs) of the images were used to identify the crystalline phases present. Energy dispersive x-ray spectroscopy (EDS) and electron energy loss spectroscopy (EELS) were used to investigate the elemental composition of the nanowires.

A typical TEM image of the nanowire showing the full diameter is shown in Figure 3(A). Analysis of multiple nanowires has indicated that the two-phase structure spans the entire length of the nanowire. A typical TEM and HRTEM image, shown in Figure 3(A,B), indicates a sharp transition of ~1-3 atomic layers between phases. The HRTEM image of Figure 3(B) also shows the highly crystalline nature of both phases.

SAED patterns were used to prove the biphasic structure of the nanowire, as shown in Figure 4. Both wurtzite phase and zinc-blende phase diffraction patterns were solved and confirmed with FFTs in the same areas where the SAEDs were taken. The FFTs and SAED patterns were checked using JAVA Electron Microscopy Simulation (JEMS[15]) and were shown to be consistent with identifications. Diffraction spots from both phases when imaged together were also noted, as shown in Figure 4(E) and indicated by the arrows. The longitudinal axis of the nanowire shown in Figure 4 is in the [110] direction for the wurtzite phase and in the [011] direction for the zinc-blende phase.

EELS and EDS spectra were taken to investigate the purity of the two crystalline phases by checking for the presence both gallium and nitrogen and the lack of oxygen. The presence of oxygen in GaN nanowires has been reported by other groups[16]. However, in order to achieve GaN nanowire growth by the method employed here, it has been found that oxygen needs to be carefully eliminated from the starting materials. Figure 5(A) shows an EELS spectrum with a strong nitrogen edge at 401eV, and the absence of an oxygen edge at 532eV, lower spectrum, and the broad gallium edge at 1115eV, upper spectrum. Figure 5(B) shows the relative intensities per energy of gallium and nitrogen, in which the gallium appears to be more intense due to its higher atomic weight. The atomic ratio of gallium to nitrogen was calculated using quantitative EDS and indicated a nearly 1:1 ratio of each element. Both EELS and EDS confirmed no detectable presence of oxygen in these nanowires. The biphasic crystallographic orientations were therefore interpreted as two phases of exclusively gallium and nitrogen.

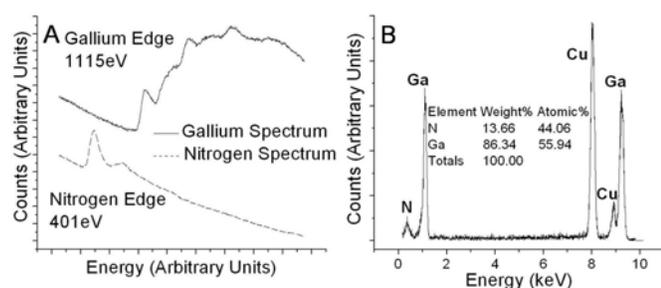

FIG 5. A. EELS spectrum showing the nitrogen edge at 401eV and the absence of an oxygen edge at 532eV, lower spectrum. EELS spectrum showing the gallium edge at 1115eV, lower spectrum, note that there is a shift in the EELS spectrum due to a floating zero peak and is within an acceptable margin of error. B. EDS spectrum showing the peaks for both gallium and nitrogen. This is indicative of the 1:1 stoichiometry of the nanowire as shown in the table. The copper peaks are present due to the copper TEM grid.

The majority of papers on GaN nanowire fabrication and structural characterization indicate that the nanowire has a pure wurtzite crystalline structure[17,18]. There has also been one report of the fabrication of GaN nanowires with a pure zinc-blende structure[19]. The possibility that our GaN nanowires were pure phase materials was investigated. The wurtzite diffraction pattern identification was checked against theoretical values for multiple diffraction planes, for both [111] zinc-blende and the [001] wurtzite zone axes as these patterns are very similar. As diffraction spots were measured further from the transmission spot, values corresponding to the [111] zinc-blende zone axis deviated substantially from ideal values, while values for the [001] wurtzite zone axis did not. The zinc-blende identification was further confirmed by the observation of [011] diffraction patterns (not shown), in addition to the [111] diffraction pattern as shown in Figure 4(D).

The basic catalyst-free nanowire growth mechanism of GaN involving the development of hexagonal crystallite platelets from an initial amorphous layer parallels the development of the nucleation layer observed in GaN thin film growth[20]. Gallium is sourced from subsequent matrix decomposition resulting first in platelet growth, then, as a kinetic equilibrium of growth and decomposition that maintains platelet size is established, in nanowire growth. Decomposition of a nucleation layer starts at ~800°C for GaN on sapphire, in good accord with the temperature for onset of GaN nanowire growth in these present experiments. Gallium atoms are highly mobile on GaN surfaces and are expected to diffuse rapidly along platelets and nanowires. Surface incorporation of Ga and ammonia to form GaN has been modeled both theoretically[21,22] and experimentally[23]. $NH_3$ appears to undergo a barrierless chemisorption on GaN, leaving $NH_2$[24,25] and H[26,27]. However, nitrogen incorporation occurs only at step edges[28]. Therefore nanowire growth occurs at the tip leading edge. We have seen no evidence for a wurtzite core growth followed by subsequent zinc-blende shell growth, and therefore conclude that, for our experimental conditions, both phases grow together along the length of the nanowire through gallium and nitrogen incorporation at the tip leading edge following a biphasic pattern initiated during early nanowire growth.

A 0.45-0.5eV band gap shift was observed in the CL measurements. An electron confinement calculation for the nanowires based on an infinite potential well model accounts for only a few meV of the observed band gap shift. Calculations based on experiment for pure wurtzite GaN nanowires have identified band gap shifts as high as 0.3eV due to compressive and tensile stresses.[29][30] Therefore the band gap shift observed in the present experiments may have more contribution from stress than from electron confinement.

To summarize, we have reported evidence for a new biphasic crystalline homostructure in GaN nanowires based on the analysis by CL and HRTEM with SAED, FFT, EDS and EELS. An inner wurtzite phase and outer zinc-blende phase crystal homostructure with a sharp phase transition of ~1-3 atomic layers have been observed. The ability to grow this biphasic GaN nanowire with longitudinal separation system of defect free zinc-blende and wurtzite phases is a new phenomenon. Biphasic nanowires based on longitudinal phase separation heterostructure[31] and now homostructure, represent a new class of electron waveguide structures with important applications in quantum transport.

1. Johnson, J.; Choi, H.J.; Knutson, K.P.; Schaller, R.D.; Yang, P.; Saykally, R.J. *Nat. Mater.*, **2002**, *1*, 106-110.
2. Huang, Y.; Duan, X.; Cui, Y.; Lieber, C. M.; *Nano Lett*. **2002**, *2*, 101-104.
3. Huang, Y.; Duan, X.; Cui, Y.; Lauhon, L. J.; Kim, K. H.; Lieber, C. M. *Sci.*, **2001**, *294*, 1313-1317.
4. Chen, X.; Xu, J.; Wang, R. M.; Yu, D.; *Adv. Mater.* **2003**, *15*, 419-421.
5. Lee, S. K.; Choi, H. J.; Pauzauskie, P.; Yang, P.; Cho, N. K.; Park, H. D.; Suh, E. K.; Lim, K. Y.; Lee, H. J. *Phys. Stat. Sol. (b)*, **2004**, *241*, 2775-2778.
6. Han, W.; Fan, S.; Li, Q.; Hu, Y. *Sci.*, **1997**, *277*, 1287-1289.

7. Duan, X. F.; Lieber, C. M. *J. Am. Chem. Soc.*, **2000**, *122*, 188-189.
8. Zhang, J.; Zhang, L. *J. Vac. Sci. Technol. B*, **2003**, *21.6.*, 2415-2419.
9. Bertness, K. A.; Roshko, A.; Sanford, N. A.; Barker, J. M.; Davydov, A. V. *J. Cryst. Growth* **2006**, *287*, 522-527.
10. He, M.; Zhou, P.; Mohammad, S. N.; Harris, G. L.; Halpern, J. B.; Jacobs, R.; Sarney, W. L.; Salamanca-Riba, L. *J. of Crys. Grow.*, **2001**, *231*, 357-365.
11. He, M.; Minus, I.; Zhou, P.; Mohammad, S. N.; Halpern, J. B.; Jacobs, R.; Sarney, W. L.; Salamanca-Riba, L.; Vispute, R. D. *Appl. Phys. Lett.*, **2000**, *77*, 3731-3733.
12. El Ahl, A. M. S.; He, M.; Peizhen, Z.; Harris, G. L.; Salamanca-Riba, L.; Felt, F.; Shaw, H. C.; Sharma, A.; Jah, M.; Lakins, D.; Steiner, T.; Mohammad, S. N. *J. of Appl. Phys.* **2003**, *94*, 7749-7756.
13. Lei, T.; Moustakas, T. D.; Graham, R. J.; He, Y.; Berkowitz, S. J. *J. Appl. Phys.,* **1992**, *71*, 4933-4943.
14. Maruska, H. P.; Tietjen, J. J. *Appl. Phys. Lett.* **1969** *15*, 327-329.
15. *http://cimewww.epfl.ch/people/stadelmann/jemsWebSite/jems.html*
16. Peng, H. Y.; Wang, N.; Zhou, X. T.; Zheng, Y. F.; Lee, C. S.; Lee, S. T.; *Chem. Phys. Lett.*, **2002**, *359*, 241-245.
17. Cheng, G. S.; Zhang, L. D.; Zhu, Y.; Fei, G. T.; Li, L.; Mo, C. M.; Mao Y. Q. *Appl. Phys. Lett.* **2006**, *75*, 2455-2457.
18. Xue, C.; Wu, Y.; Zhuang, H.; Tian, D.; Liu, Y.; Zhang, X.; Ai, Y.; Sun, L.; Wang, F. *Physica E*, **2005**, *30*, 179-181.
19. Han, W. Q.; Zettl, A. *Appl. Phys. Lett.* **2002**, *81*, 5051-5053.
20. Wikenden, A. E.; Wickenden D. K.; Kietenmacher, T. J. *J. of Appl. Phys.* **1994**, *75*, 5367-5371.
21. Fritsch, J.; Sankey, O. F.; Schmidt, K. E.; Page, J. B. *Surf. Sci.* **1999**, *428*, 298-303.
22. Northrup, J. E.; Di Felice, R.; Neugebauer, J. *Phys. Rev. B* **1997**, *56*, R4325-R4328.
23. Brown, J. S.; Koblmueller, G.; Wu, F.; Averbeck, R.; Riechert, H.; Speck, J. S. *J. Appl. Phys.* **2006**, *99*, 074902.
24. Bermudez, V. M. *Chem. Phys. Lett.* **2000**, *317*, 290-295.
25. Pignedoli, C. A.; Di Felice, R.; Bertoni, C. M. *Phys. Rev. B* **2001**, *64*, 113301.
26. Gherasoiu, I.; Nikishin, S. and Temkin, H. *J. Appl. Phys.* **2005**, *98*, 053518.
27. McGinnis, A. J.; Thomson, D.; Davis, R. F.; Chen, E.; Michel, A.; Lamb, H. H. *Surf. Sci.* **2001**, *494*, 28-42.
28. Held, R.; Ishaug, B. E.; Parkhomovsky, A.; Dabiran, A. M.; Cohen, P. I. *J. Appl. Phys.* **2000**, *87*, 1219-1226.
29. Seo, H. W.; Bae, S. Y.; Park, J.; Yang, H. N.; Park, K. S.; Kim, S. *J. Chem. Phys.* **2002**, *116*, 9492-9499.
30. Kuykendall, T.; Pauzauskie, P. J.; Zhang, Y.; Goldberger, J.; Sirbuly, D.; Denlinger, J.; Yang, P. *Nat. Mater.* **2004**, *3*, 524-528.
31. Lauhon, L. J.; Gudiksen, M. S.; Wang, D.; Lieber, C. M. *Nature*, **2002**, *420*, 57-61.